\documentclass[amsmath,amssymb,prd,preprintnumbers,showpacs]{revtex4}
\newcommand{\overbar}[1]{\mkern 1.5mu\overline{\mkern-1.5mu#1\mkern-1.5mu}\mkern 1.5mu}
\usepackage{bbold}
\usepackage[cmtip,arrow]{xy}
\usepackage{pb-diagram,pb-xy}
\usepackage{slashed}
\usepackage{graphicx}
\usepackage{amsmath}
\usepackage{amssymb}
\usepackage{float}
\usepackage{amsopn}
\usepackage{fancyhdr}
\usepackage[yyyymmdd,hhmmss]{datetime}
\newcommand{\be}{\begin{equation}}
\newcommand{\ee}{\end{equation}}
\newcommand{\en}{\end{equation}}
\newcommand{\ba}{\begin{eqnarray}}
\newcommand{\ea}{\end{eqnarray}}
\newcommand{\bea}{\begin{eqnarray}}
\newcommand{\eea}{\end{eqnarray}}

\def \be {\begin{equation}}
\def \ee {\end{equation}}
\def \bea {\begin{eqnarray}}
\def \eea {\end{eqnarray}}

\begin{document}
\title{Perturbative unitarity and higher-order Lorentz symmetry breaking }
\author{ Leonardo Balart$^{1}$}
\email[Electronic mail: ]{leonardo.balart@ufrontera.cl}
\author{ Carlos M. Reyes$^{2}$}
\email[Electronic mail: ]{creyes@ubiobio.cl}
\author{Sebastian Ossandon$^{3}$ }
\email[Electronic mail: ]{sebastian.ossandon@pucv.cl}
\author{ Camilo Reyes$^{4}$}
\email[Electronic mail: ]{creyesm70@gmail.com}
\affiliation{$^{1}$ Departamento de Ciencias F\'isicas, Facultad de Ingenier\'ia y Ciencias, Universidad 
de La Frontera, Casilla 54-D, Temuco, Chile}
\affiliation{$^{2}$ Grupo de Cosmlog\'{\i}a y Part\'{\i}culas Elementales, Departamento de Ciencias B\'{a}sicas, Universidad del B\'{\i}o B\'{\i}o,
Casilla 447, Chill\'{a}n, Chile}
\affiliation{$^{3}$ Instituto de Matem\'aticas, Pontificia Universidad 
Cat\'olica de Valpara\'iso, Casilla 4059, Valpara\'iso, Chile}
\affiliation{$^{4}$  Facultad de Ingenier\'{\i}a, Ciencia y Tecnolog\'{\i}a, 
Universidad Bernardo O'Higgins, Avenida Viel 1497, C\'odigo Postal 8370993, Santiago, Chile}

\begin{abstract}
We study perturbative unitarity in the scalar sector of the 
Myers-Pospelov model. The model introduces a preferred four-vector $n$ which 
breaks Lorentz symmetry and couples to a five-dimension operator. When the preferred four-vector
is chosen in the pure timelike or lightlike direction, the model becomes a higher time derivative theory, leading to 
a cubic dispersion relation.
Two of the poles are shown to be perturbatively connected to the standard ones, while a third pole, 
which we call the Lee-Wick-like pole, is associated to a negative metric, in Hilbert space, threatening
the preservation of unitarity.
The pure spacelike case is a normal theory in the sense that 
it has only two solutions both being small perturbations over the standard ones.
We analyze perturbative unitarity for purely spacelike and timelike cases
using
 the optical theorem and considering a quartic self-interaction term. 
By computing discontinuities in the loop diagram, we arrive at a pinching condition 
which determines the propagation 
of particles and Lee-Wick-like particles through the cut. We find that the contribution for Lee-Wick-like particles vanishes for any
external momenta, leaving only the contribution of particles,
 thus preserving one-loop unitarity in both cases.
\end{abstract}
\maketitle

\section{Introduction}
The breakdown of Lorentz symmetry at the Planck mass 
scale $m_P=10^{19}$ GeV
has been intensively studied in the last two decades.
Many efforts have been put forward in order
to provide experimental input for these 
quantum gravity effects. 
The detection of such possible new physics, however, has been
 very challenging principally because
  the high scale imposes a strong suppression.
In particular, at present-time colliders with attainable energies of $m\sim13$ TeV,
these effects may be suppressed by some power of $m/m_P \sim 10^{-15}$, which is very small.
Moreover, for experiments using 
 the highest-energy cosmic rays observed, they are
 about
eight orders of magnitude below the Planck mass. $CPT$ and Lorentz symmetry 
departures have received motivation from various sources, specially 
from attempts to construct a 
quantum gravity theory~\cite{QGV}.

The effective approach, encoding the
 high scale $\Lambda$, has shown to be
 a powerful method to explore such departures. 
One advantage is that it allows us to include the most general form of Lorentz invariance violation, 
without resorting to a particular theory or method of calculation to get down to a 
low-energy model.
Many of these studies have been given 
within the effective framework of the standard model extension (SME)~\cite{SME}.
The SME encompasses all the possible effective terms describing
Lorentz symmetry violation in matter and gravity sectors.
It is implemented through constant coefficients
which couple to operators of renormalizable mass dimension in the minimal sector
and to higher-order mass dimension 
operators in the nonminimal sector~\cite{SME-nonminimal}. 
The coefficients are believed to arise as expectation values of 
tensor fields possibly from spontaneous Lorentz violation in a more fundamental theory. 
The strong bounds on these parameters in the minimal sector has prompted
the exploration at higher energies using higher-order 
operators~\cite{MP}. Limits on the
Lorentz violating coefficients of nonrenormalizable operators
have been obtained from astrophysical 
observations~\cite{MP_Limits} and synchrotron radiation~\cite{syn};
see also~\cite{comp}. Recently, extensions 
with higher-order couplings have also
been proposed~\cite{H-O-C}.

One theoretical advantage of introducing higher-order 
operators is that ultraviolet divergencies of conventional 
quantum field theories
can be 
softened~\cite{Rad-Corr,dim5}. However, as is well 
known, in many cases this comes with the 
 appearance of an indefinite-metric Hilbert space,
   leading to 
a possible loss of conservation of probability 
or nonunitarity of the $S$ matrix~\cite{PU}. 
Many of the problems have been analyzed and resolved in the framework 
developed by Lee and Wick \cite{Lee-Wick}, in which the asymptotic space is 
restricted to contain only stable particles with positive metric.
Further studies to deal 
with amplitudes in a covariant fashion and their nonanalytic pinching within an \emph{ad hoc} 
prescription,
were developed in~\cite{Cut}.
The indefinite metric approach has also been used to improve the 
hierarchy problem in the scalar sector of the standard model~\cite{G}.
Recently, it has been shown that Lee-Wick theories can 
be interpreted as nonanalytic Wick rotated  Euclidean theories~\cite{Ans-Piva}. Here, we study 
unitarity in a Lorentz violating model with higher-order operators in light of the 
Lee-Wick studies~\cite{unit-LW}. The class of higher time derivative field theories that
we consider
 extends the notion of Lee-Wick theory due to the explicit noncovariance,
 which may be reflected by the absence of complex conjugate ghost poles.
 Previous studies for tree level unitarity have been given in~\cite{unit}.

Another focus, recently discussed in~\cite{Ext-leg,dens}, concerns the effect of
 Lorentz violating radiative corrections in tree level physics.
It has been shown that external leg physics gets
modified due to the appearance of observer Lorentz scalars in the 
spectral density function~\cite{dens}. A particular model of the SME has been analyzed
and the corresponding modification for asymptotic fields has been found~\cite{Ext-leg}.
Extensions to include higher-order Lorentz violation have been given in~\cite{ren}.

The organization of this paper is as follows.
In Sec.~\ref{sectionII}, we introduce the scalar Myers-Pospelov 
model with dimension-five operators.
We found the dispersion relations for the
purely spacelike, purely timelike, and lightlike cases.
We analyze the solutions in the three cases
and found that for certain values of space momenta, 
some solutions become complex, making the poles of the propagators move to the complex energy plane.
In all of the cases, we identify perturbative solutions and those belonging to the Lee-Wick-like class
 with an associated negative metric.
In Sec.~\ref{sectionIII}, we prove the conservation of unitarity for the purely spacelike and timelike cases
 at one-loop order
using the optical theorem and focusing on a quartic interaction 
term. Finally, in Sec.~\ref{con}, we give 
our final remarks and conclusions.
\section{Effective model}\label{sectionII}
Our model is based on the Lorentz violating extension 
 Myers-Pospelov Lagrangian density in the scalar sector~\cite{MP}:
\begin{eqnarray}\label{Lag-MP}
\mathcal L&=&\partial_{\mu} \Phi^{*} \partial^{\mu}\Phi  
-m^2\Phi^{*} \Phi+i g  \Phi^{*} (n\cdot \partial)^3 \Phi +\mathcal L_{\text{int}} \,,
\end{eqnarray}
where 
\begin{eqnarray}
\mathcal L_{\text{int}}&=&\frac{\lambda}{4!}  (\Phi^{*}\Phi)^2  \,.
\end{eqnarray}
Here $g$ is a Planck mass suppressed constant 
and $n$ a preferred four-vector
which characterizes the type of Lorentz violation. When $n$
has a temporal component,
the theory belongs to a class of theory better known as 
higher time derivative theory.
As we will show further in this case the theory 
displays an additional degree of freedom.

To begin with, let us consider a general preferred four-vector $n$ with a free equation of motion ($\lambda=0$):
 \begin{eqnarray}
\left( \Box +m^2-ig\left(n\cdot \partial\right)^3 \right) \Phi=0\,.
\end{eqnarray}
Using  the plane wave ansatz
$\Phi(x) \sim \int dp \,\Phi(p)e^{-ipx}$ yields the dispersion relation 
\begin{eqnarray}\label{spacelike}
p^2 -m^2-g(n\cdot p)^3 =0\,.
\end{eqnarray}
Now we can specialize to the different cases. Let us begin 
with a pure spacelike four-vector 
$n=(0,\vec n)$, where the dispersion relation takes the form
\begin{eqnarray}\label{disp-rel-spacelike}
p_0^2-E_p^2+g (\vec n \cdot \vec p)^3=0\,,
\end{eqnarray}
where $E_p=\sqrt{\vec p^2+m^2}$. The solutions are
$p_0=\pm \omega_s $ with 
\begin{eqnarray}\label{sol_scalar}
\omega_s=   \sqrt{E_p^2-g  (\vec n \cdot \vec p)^3   }  \,.
\end{eqnarray}
Without loss of generality we take $g>0$ and define the function 
\begin{eqnarray}\label{func}
f(|\vec p|)&=& |\vec p|^2+m^2-a |\vec p|^3  \,,
\end{eqnarray}
with
\begin{eqnarray}
 a(\theta) =g |\vec n|^3 \cos^3 \theta 
\end{eqnarray}
where $\theta$ is the angle between $\vec n$ and $\vec p$.

When $0\le \theta <\pi/2$, and so $a>0$, some solutions 
become complex at higher momenta than 
 $| \vec p|> \mathcal P$, where we define
\begin{eqnarray}
\mathcal P
   &=&  \frac{1}{3a}\left(  1+\left (\frac{1+i\sqrt{3}}{2}\right)Q^{-1/3} +\left (\frac{1-i\sqrt{3}}{2}\right)Q^{1/3}    \right)\,,
\end{eqnarray}
with
\begin{eqnarray}
 Q&=&\frac{1}{2}  \left(   b+\sqrt{b^2-4} \right)\,,\nonumber  \\ b&=&-2-27a^2m^2\,.
\end{eqnarray}
One can show that for $gm\ll |\vec n|^{-3}\cos ^{-3}\theta$, the approximation gives
\begin{eqnarray}
 \mathcal P\approx \frac{1}{g |\vec n|^3 \cos^3 \theta}+ gm^2 |\vec n|^3 \cos^3 \theta   \,,
\end{eqnarray}
which indeed is very high for any angle $\theta$ in the interval.
In contrast, at directions $\pi/2  < \theta\le \pi$, the solutions $\omega_s$ are 
always real. For the particular value at $\theta =\pi/2$, we have 
a blind direction at which we  
recover the usual 
dispersion relation.

In our concordant frame which follows from the condition imposed on $g$,
there may be instabilities due to complex solutions that arise for higher values than $\mathcal P$, but also 
 instabilities related to spacelike states~\cite{K-R}. This is true even 
 imposing the cutoff 
 $\mathcal P$, since then highly boosted frames with real momenta can 
 produce negative energies.
As an example, consider  
 $n=(n_0,0,0,n_3)$ and the corresponding 
spacelike state $p=(0,0,0,p_{3}^{\pm})$,
\begin{eqnarray}
p_{3}^{\pm}=\frac{gn_3}{2}\pm \sqrt{\left(\frac{gn_3}{2}\right)^2-m^2  } \,,
\end{eqnarray}
which is a solution of the dispersion relation \eqref{disp-rel-spacelike}. 

A more general analysis
follows by considering the velocity group $v_g$, 
\begin{eqnarray}
v_g=\frac{2|\vec p |+3a   |\vec p |^2}{2\sqrt{|\vec p|^2+m^2+a|\vec p|^3   }}\,,
\end{eqnarray}
where we expect instabilities (provided interactions are turned on) and small 
deviations from microcausality as seen in the limit $v_g\to \infty$
for $|\vec p| \to \infty$, see ~\cite{K-R}.
Recently, it has been shown that an extended Hamiltonian formalism allows us
to implement
a consistent canonical quantization for 
Lorentz violating theories containing spacelike states~\cite{Neg-En}.

It is not difficult to find the propagator 
\begin{eqnarray}\label{spacelike-propagator}
i\Delta(p_0,\vec p, \epsilon)&=&\frac{i}{(p_0-\omega+i\epsilon) 
 (p_0+\omega -i\epsilon) }  \,,
\end{eqnarray}
where the location of poles is the standard one.

\begin{figure}
\centering
\includegraphics[width=0.45\textwidth]{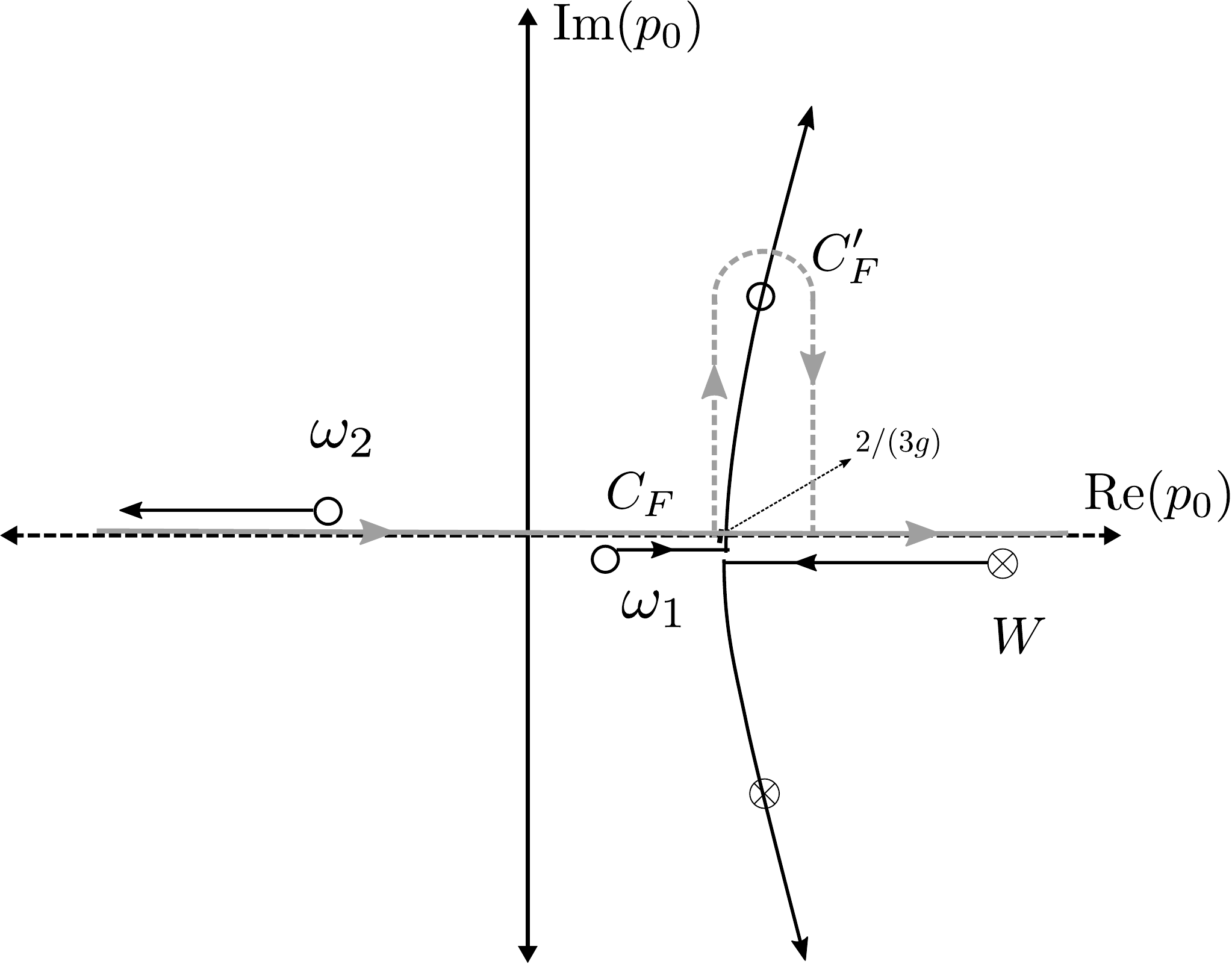}
\caption{\label{Fig1} The path of integration $C_F$ when all the poles are real and 
the deformed path $C^{\prime}_F$ when $\omega_1$ and $W$ become complex.}
\end{figure}

Now we continue with a
purely timelike four-vector $n=(1,0,0,0)$.  It yields the dispersion relations 
\begin{eqnarray}\label{alg-eq}
p_0^2-E_p^2-gp_0^3=0\,.
\end{eqnarray}
Solving \eqref{alg-eq} we find the exact three solutions 
\begin{eqnarray}\label{solu}
   \omega_1&=&\frac{1}{3g} \left(1+  \xi^{-1/3}  z _0
   + \xi^{1/3} z_0^{*}  \right)     \,,  \nonumber \\ \omega_2
   &=&  \frac{1}{3g}\left(  1-\xi^{-1/3} -\xi^{1/3}    \right)\,,
  \\ 
W&=&\frac{1}{3g}\left(1+   \xi^{-1/3}  z_0^{*} 
+ \xi^{1/3} z_0 \right)\,, \nonumber
\end{eqnarray}
where we have introduced 
 $z_0=e^{-\frac{i \pi}{3}}$
and defined the expressions
\begin{eqnarray}\label{def-xi}
\xi&=&\frac{1}{2}(  \beta + \sqrt{\beta^2-4}) \,, \nonumber \\ 
 \beta&=&-2 + 27g^2 E_p^2\,.
\end{eqnarray}
In complex energy-plane in terms 
 of momenta $|\vec p|$, the solutions move according to~Fig.\ref{Fig1}.
In an analogous way, the solutions for the field $\Phi^*$ 
are obtained by the replacement $g\to -g$. 

\begin{figure}
\centering
\includegraphics[width=0.4\textwidth]{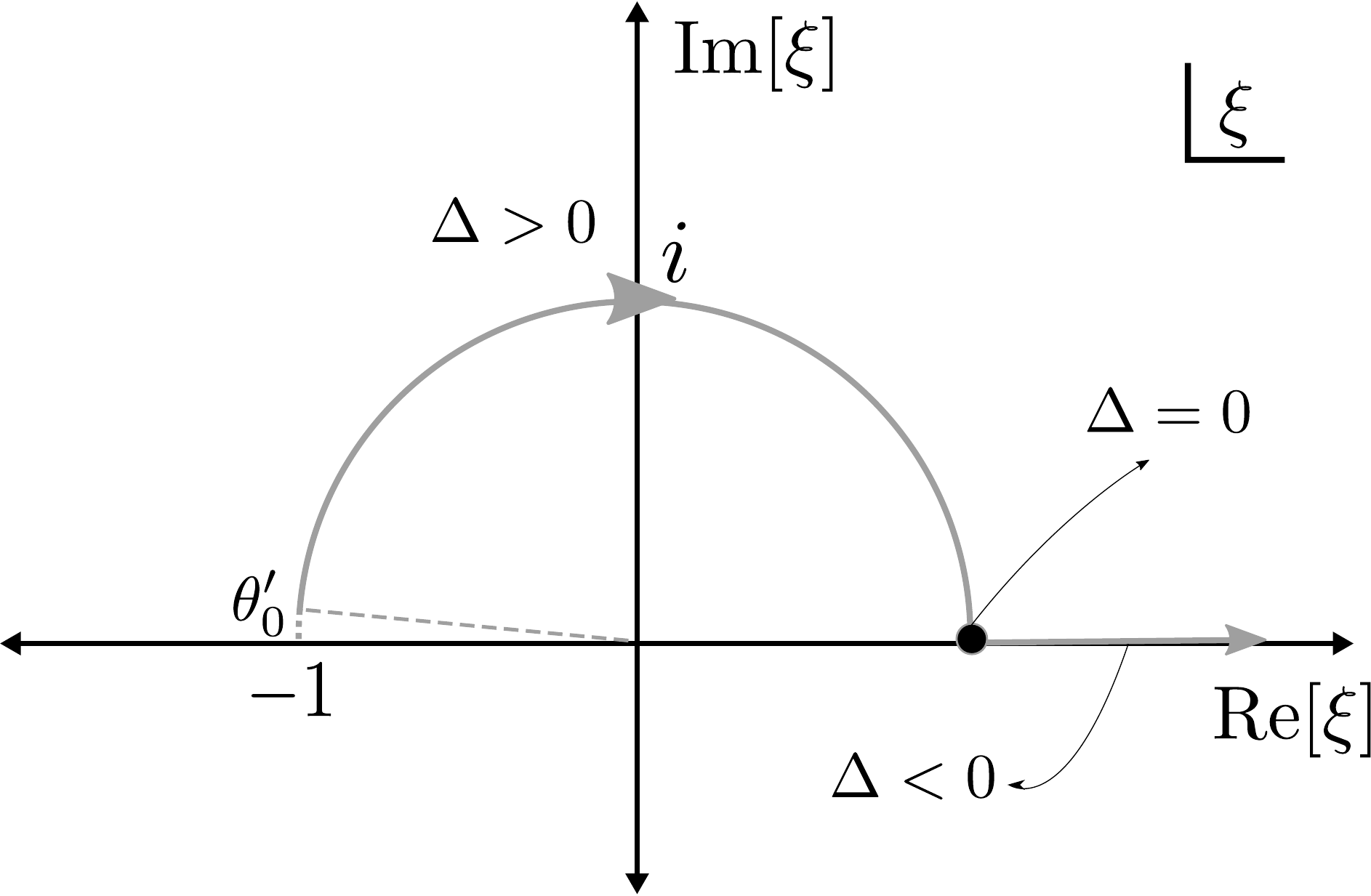}
\caption{\label{Fig2} The energy function $\xi$ in terms of $\Delta$.}
\end{figure}

The solutions~\eqref{solu} can be classified according to the sign of the
discriminant $\Delta=E_p^2(2-\beta)$, which leads to 
 the following three cases:
 
\begin{enumerate}
\item[i)] When $\Delta>0$, all solutions are real 
($\beta<2$) and $\xi$ is a complex number that moves in the clockwise direction on a semicircle of 
unit norm, starting at the angle $\theta^{\prime}_{0}=gm \sqrt{27}$; see Fig.~\ref{Fig2}. 
\item[ii)] When $\Delta=0$, which we call the critical value ($\beta=2$), the two solutions $\omega_1$ and $W$ 
collapse at $\frac{2}{3g}$, and $\omega_2=-\frac{1}{3g}$, which may be seen using $\xi=1$ in Eq.~\eqref{solu}.
\item[iii)] When $\Delta<0$, the two solutions $\omega_1$ and $W$ become complex 
 conjugate pairs, i.e., $\omega_1=W^*$ $(\beta>2)$,
while the solution $\omega_2$ remains real. We have that $\xi$ is a real number
  larger than $1$, see Fig.~\ref{Fig2}. 
\end{enumerate}

In order to characterize the poles, let us consider the asymptotic expansion
for the limit $g\to 0$
\begin{eqnarray}
\omega_1 &=& E_p + \frac{gE_p^2}{2}+ \frac{5g^2E_p^3}{8}+\mathcal O(g^3)\,,  \nonumber
 \\ \omega_2 &=& - E_p +  \frac{gE_p^2}{2}  - \frac{5g^2E_p^3}{8}+\mathcal O(g^3) \,,     \\
W& =& \frac 1 g -gE_p^2-2g^3E_p^4+\mathcal O(g^5)   \nonumber \,,
\end{eqnarray}
valid for $E_p \ll 1$. 
As usual, one can associate $\omega_1$ and $\omega_2$ to a
particle and antiparticle, respectively, while 
$W$ to an additional particle which we call the Lee-Wick-like particle.
In this sense, one may
regard the higher time 
derivative theory with an indefinite metric
as a Lee-Wick-like extension given that the poles
$W$ and $\omega_1$ only become
complex conjugate pairs in a
certain range of
energies, as we will see below.

The Feynman propagator can be defined as
\begin{eqnarray}\label{prop}
i\Delta(p_0,\vec p, \epsilon)&=&\frac{i}{(p_0-\omega_1+i\epsilon) 
 (p_0-\omega_2-i\epsilon) }  \nonumber   \\  &\times&\frac{1}{-g(p_0-W+i\epsilon)} \,,
\end{eqnarray}
with the negative pole $\omega_2$ located in the 
second quadrant and positive poles $\omega_1$ and $W$ in the fourth; see~Fig.~\ref{Fig1}.
 That is, $\omega_1$ and 
$W$ ($\Delta>0$) lie below the path of integration $C_F$
and $\omega_2$ above. The poles $\omega_1$ and $W$
move in the opposite direction in the real axis collapsing 
at $\frac{2}{3g}$ ($\Delta=0$), while 
$\omega_2$ always moves to the left in the real axis.
For energies
($\Delta<0$) the equivalent path $C^{\prime}_F$ 
rounds the complex solution $\omega_1$ from above.
 This prescription enjoys the 
desirable property to recover the standard
 position of the perturbative
 poles in the limit $g\to 0$ and to be connected to the 
 Euclidean theory through a Wick rotation.
 In Fig.~\ref{Fig1} the circles denote the perturbative poles 
and the encircled crosses the Lee-Wick-like pole. 

The analysis for a lightlike four-vector $n=(1,0,0,1)$ is very 
similar to the
 previous case. By considering the dispersion relation 
 \begin{eqnarray}
p_0^2-E_p^2-g(p_0-p_3)^3=0 \,,
\end{eqnarray}
 the solutions are 
 \begin{eqnarray}\label{light-solu}
   \gamma_1&=&\frac{1}{3g} \left( 1+3gp_3+ (1+6gp_3) \eta^{-1/3}  z _0
   + \eta^{1/3} z_0^{*}  \right)     \,,  \nonumber \\ \gamma_2
   &=&  \frac{1}{3g}\left(  1+3gp_3- (1+6gp_3) \eta^{-1/3} -\eta^{1/3}    \right)\,,
  \\ 
\gamma_3&=&\frac{1}{3g}\left(1+3gp_3+  (1+6gp_3) \eta^{-1/3}  z_0^{*} 
+ \eta^{1/3} z_0 \right)\,, \nonumber
\end{eqnarray}
where again 
 $z_0=e^{-\frac{i \pi}{3}}$
and we have defined the expressions
\begin{eqnarray}
\eta&=&\frac{1}{2}\left(  \delta + \sqrt{ \delta^2-4  (1+6gp_3)}\right) \,, \nonumber \\ 
 \delta&=&-2 + 27g^2 E_p^2-18gp_3-27g^2p_3^2\,.
\end{eqnarray}
As before, considering the limit $g\to 0$, we 
identify the solution $\gamma_3$ with the propagation of a Lee-Wick-like state.
\section{Perturbative unitarity}\label{sectionIII}
\begin{figure}[H]
\centering
\includegraphics[width=0.28\textwidth]{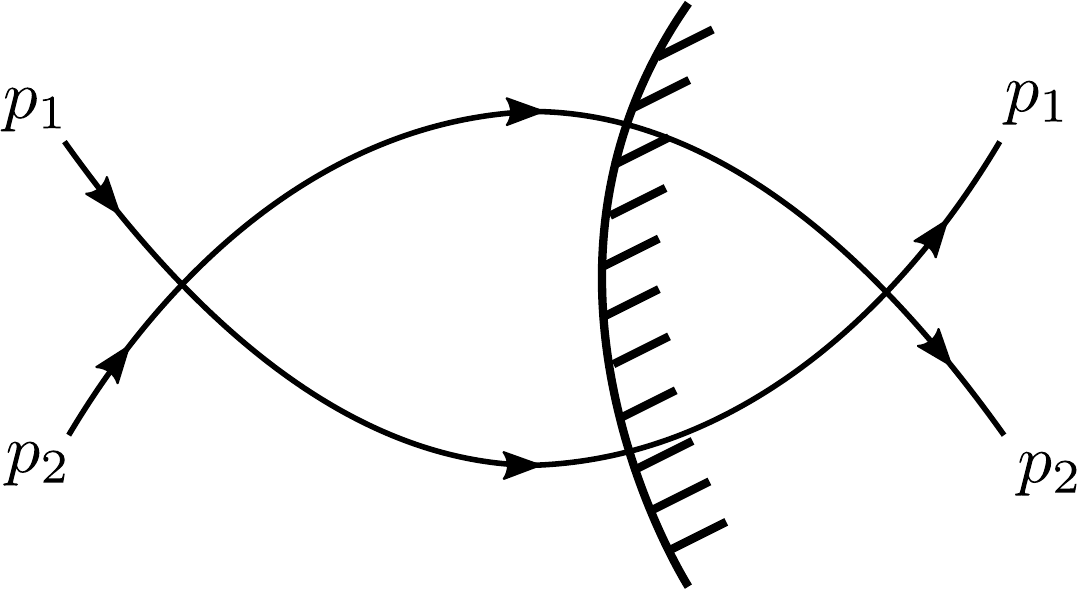}
\caption{\label{Fig3} The forward scattering of 
a particle and antiparticle with 
momenta $p_1$ and $p_2$, respectively.}
\end{figure}

In this section we study perturbative unitarity at one-loop
order for the graph shown in Fig.~\ref{Fig3} by considering 
a purely timelike and spacelike preferred four-vector.
We write the loop amplitude 
\begin{eqnarray}\label{amplitude}
i\mathcal M(p)&=&(i\lambda)^2 \int \frac{d^4q}{(2\pi)^4}
 \Delta(q_0,\vec q,\epsilon)   \Delta(q_0-p_0,\vec q- \vec p ,\epsilon)  
\end{eqnarray}
in terms of a generic external 
momenta $p=(p_0,\vec p)$ such that $p=p_1+p_2$. Here
energy flows through
the cut towards the shaded region as shown in Fig~\ref{Fig3}.
\subsection{Purely spacelike $n$}\label{subsectionI}
Consider a spacelike four-vector 
$n=(0,0,0,\vec n)$ in the amplitude Eq.~\eqref{amplitude} and with the propagator~\eqref{spacelike-propagator}
\begin{eqnarray}
i\mathcal M(p)&=&-\lambda^2 \int \frac{d^4q}{(2\pi)^4}
\frac{i}{(q_0-\omega_s+i\epsilon) 
 (q_0+\omega_s -i\epsilon) }   \nonumber \\
 &\times&\frac{i}{(q_0-p_0-\Omega_s+i\epsilon) 
 (q_0-p_0+\Omega_s -i\epsilon) }  \,.
\end{eqnarray}
The notation is 
\begin{eqnarray}
\Omega_{s}&=&\sqrt{(\vec p-\vec q)^2+m^2-g 
 (\vec n \cdot (\vec p-\vec q))^3   } \,, \\ \omega_s&=&   \sqrt{\vec q^2+m^2-g  (\vec n \cdot \vec p)^3   }  \,,
\end{eqnarray}
with the last term defined in Eq.~\eqref{sol_scalar}.

The integration over $q_0$ is performed with the method of residues 
and we close the 
contour from below, enclosing the poles $\omega_s-i\epsilon$ and $p_0+\Omega_s-i\epsilon$.
After summing the two contributions we have
\begin{eqnarray}
i\mathcal M(p)&=&i\lambda^2 \int \frac{d^3q}{(2\pi)^3}
\frac{(\omega_s+\Omega_s-2i\epsilon)}{ 2(\omega_s-i\epsilon) (\Omega_s-i\epsilon)  (\omega_s +\Omega_s -p_0   -2i\epsilon) 
 (\omega_s +\Omega_s +p_0   -2i\epsilon) }   \,.
\end{eqnarray}
Next, we set $\epsilon=0$, where it does not 
 affect the computation of the discontinuity, which follows from the identity
\begin{eqnarray}
\frac{1}{x\pm i\epsilon}=\mathcal{P}\left(\frac{1}{x}\right)\mp i\pi \delta(x)\,.
\end{eqnarray}
In this way we arrive at
\begin{eqnarray}
\text{Disc}\mathcal M(p)&=&i\lambda^2  \int \frac{d^3q}{(2\pi)^3}(2\pi) \left[
\frac{\delta(p_0-\omega_s -\Omega_s    ) }{ 4\omega_s \Omega_s   
  } + \frac{\delta(p_0+\omega_s +\Omega_s   ) }{ 4\omega_s \Omega_s   
  }  \right]   \,.
\end{eqnarray}
We introduce the four-vectors $q_1=(q_{01}, \vec q)$ and $q_2=(q_{02}, \vec p-\vec q)$ with $q_{01}=\omega_s$
and $q_{02}=\Omega_s$. With this, we first rewrite 
\begin{eqnarray}
\text{Disc}\mathcal M(p)&=&i\lambda^2  \int \frac{d^3q}{(2\pi)^3}(2\pi)   \int dq_{01}\int dq_{02} \delta(p_0-q_{01} -q_{02}    ) \nonumber \\ &\times& \left[
\frac{\delta(q_{01}-\omega_s    )  \delta(q_{02}-\Omega_s    ) }{ 4\omega_s \Omega_s   
  } + \frac{\delta(q_{01}-\omega_s    )  \delta(q_{02}-\Omega_s    ) }{ 4\omega_s \Omega_s   
  }  \right]   \,,
\end{eqnarray}
and then by using
\begin{eqnarray}
\int \frac{d^3q}{(2\pi)^3}  =\int \frac{d^3 q_1}{(2\pi)^3} \int \frac{d^3q_2}{(2\pi)^3} (2\pi)^3 \delta^{(3)}(\vec p-\vec q_1-\vec q_2)\,,
\end{eqnarray}
we transform the integral into
\begin{eqnarray}
\text{Disc}\mathcal M(p)&=&i\lambda^2 \int \frac{d^4q_1 }{(2\pi)^4}   \int \frac{d^4q_2 }{(2\pi)^4} 
   (2\pi)^4 \delta^{(4)}(p-q_1 -q_2   ) 
\left[ \frac{  (2\pi) \delta(q_{01}-\omega_s)   (2\pi) \delta(q_{02}-\Omega_s   )    }{ 2 \omega_s 2 \Omega_s   
  } \right. \nonumber  \\  &+& \left. \frac{  (2\pi) \delta(q_{01}+\omega_s)   (2\pi) \delta(q_{02}+\Omega_s   )    }{ 2 \omega_s 2 \Omega_s   
  }  \right]    \,.
\end{eqnarray}
Finally, using the identity $\text{Disc}\mathcal M=2i \text{Im} M$, we have
\begin{eqnarray}\label{spacelike-unit}
2 \text{Im}\mathcal M(p)&=&\lambda^2 \int 
\frac{d^4q_1}{(2\pi)^4}   \int \frac{d^4q_2}{(2\pi)^4} (2\pi)^4  \delta^{(4)}(p-q_1-q_2) \delta(q_1^2-m^2-g(\vec n \cdot \vec q_1)^3)
 \delta(q_2^2-m^2-g(\vec n \cdot \vec q_2)^3)\nonumber  \\ &\times& \left[   \theta(q_{01}     \theta (q_{02}))  +\theta(-q_{01} )    \theta (-q_{02})
 \right] \,.
\end{eqnarray}
From the cut diagram of the right, we identify the sum over intermediate states,
the conservation of momenta coded in the first delta and the two
 propagators put on shell through the deltas of the dispersion relation.
We may further simplify the result by considering the routing where energy flows with positive $q_{01}$ and $q_{02}$.
Finally, we have that the optical theorem is satisfied in our process at the one-loop level.
\subsection{Purely timelike $n$}\label{subsectionI}
From the previous sections, we 
have seen that
a Lee-Wick-like particle arises
when $n$ is chosen in the purely timelike direction.
In order to study unitarity, we follow the 
Lee-Wick prescription in which 
 only positive-norm states are regarded as stable, so removing
 from the asymptotic space the Lee-Wick-like particles~\cite{Lee-Wick}. The prescription is far from being trivial since 
Lee-Wick-like states may arise within the loops, spoiling any attempt to preserve unitarity.
Hence, as a general statement one can say that if unitarity is to be 
 conserved, no 
 Lee-Wick-like states should propagate through the cut. 

\begin{figure}
\centering
\includegraphics[width=0.45\textwidth]{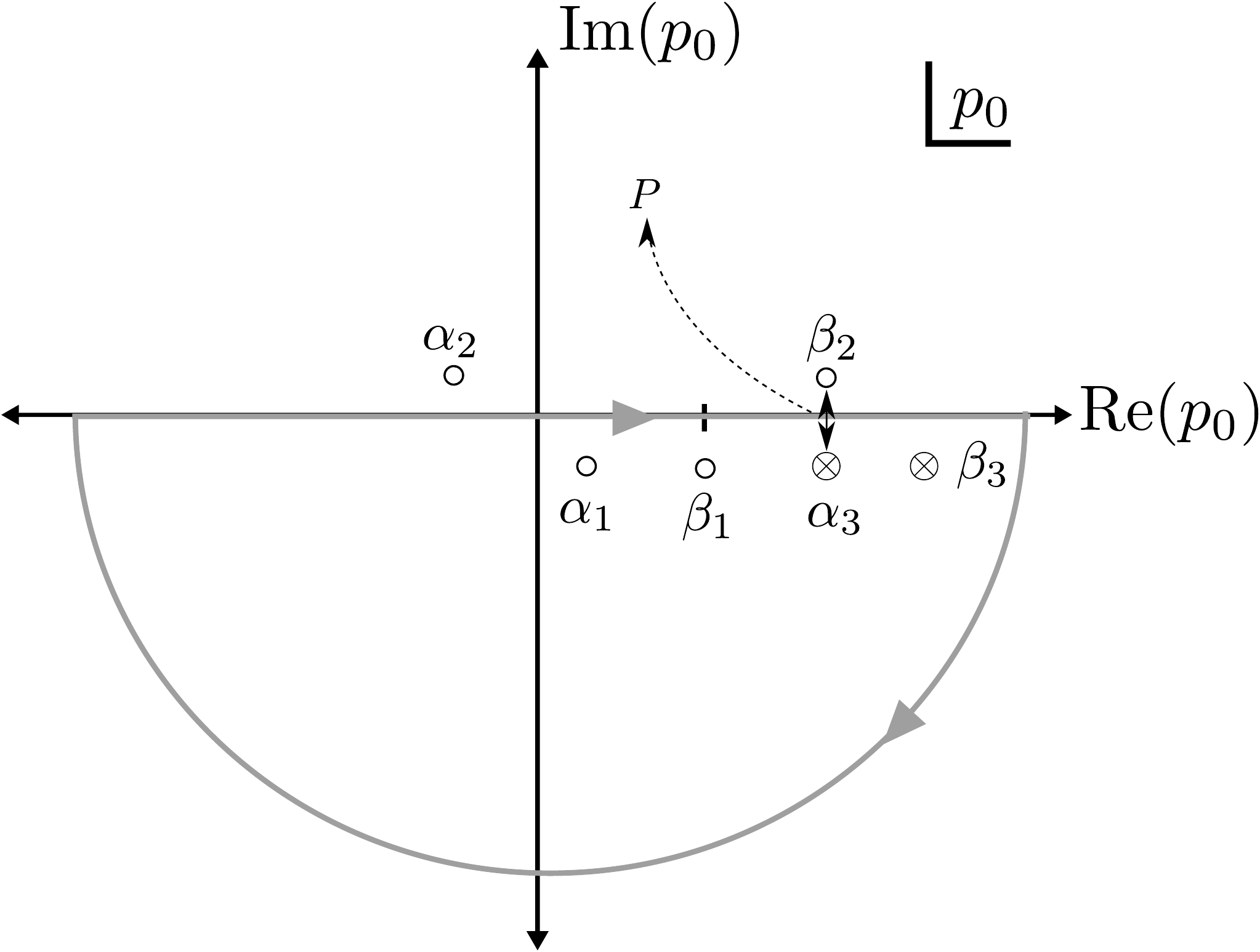}
\caption{\label{Fig4} The relevant poles $\alpha_1$, $\beta_1$, $\alpha_3$, $\beta_3$ 
and the pinching point $P$ for the collapse of $\beta_2$ and $\alpha_3$ on the contour of integration.}
\end{figure}

The amplitude is written with the propagator of Eq.~\eqref{prop} 
\begin{eqnarray}
i\mathcal M(p)&=& \lambda^2   \int \frac{ d^{4}{ q}}{(2\pi)^4} \prod_{i=1}^2
\frac{1}{ (q_0^{(i)}-\omega_1^{(i)}+i\epsilon) (q_0^{(i)}-{\omega_2^{(i)}}-i\epsilon)  } 
 \nonumber \\ &\times& 
\frac{1}{   -g (q_0^{(i)}-W^{(i)}+i\epsilon)    }    \,,
\end{eqnarray}
and with the new notation where $q_0^{(1)}=q_0$ and $q_0^{(2)}=q_0-p_0$, together with
\begin{eqnarray}
\omega_1^{(1)}&=&\omega_1(\vec q)\,,   \qquad \omega_1^{(2)}
=\omega_1(\vec p-\vec q) \,, \nonumber \\
\omega_2^{(1)}&=&\omega_2(\vec q)\,, \qquad \omega_2^{(2)}
=\omega_2(\vec p-\vec q)\,, \nonumber \\  
W^{(1)}&=&W(\vec q)\,, \qquad W^{(2)}=W(\vec p-\vec q).
\end{eqnarray}
The first propagator has poles at 
\begin{eqnarray}
\alpha_1&=&\omega_1 (\vec q) -i\epsilon  \,,  \nonumber   \\
\alpha_2&=&\omega_2  (\vec q)+ i\epsilon  \,,   \nonumber \\
 \alpha_3&=&W (\vec q)- i\epsilon   \,,    
\end{eqnarray}
 and the second propagator, which depends on the external 
 momenta, has poles at
 \begin{eqnarray}
\beta_1&=&p_0 + \omega_1(\vec q-\vec p)  -i\epsilon   \,,  \nonumber        \\
\beta_2&=& p_0 + \omega_2 (\vec q-\vec p)  +i\epsilon \,,   \nonumber\\
 \beta_3&=& p_0 + W (\vec q-\vec p)  -i\epsilon  \,,       
\end{eqnarray}
they are depicted in Fig.~\ref{Fig4}.

Let us perform the integral in $q_0$
using the residue theorem and
closing the contour of $q_0$ in the lower half plane. In this way, we 
enclose the poles 
$\alpha_1$, $\alpha_3$, $\beta_1$, and $\beta_3$ to obtain
\begin{eqnarray}
i\mathcal M(p)&=&-i \lambda^2 \int \frac{d^{3}
{\vec q}}{(2\pi)^{3}}  \Big( \text{Res}(\alpha_1) +\text{Res}(\alpha_3)  \nonumber \\
&+&\text{Res}(\beta_1) +\text{Res}(\beta_3)  \Big) \,,
\end{eqnarray}
where the corresponding residues are
\begin{eqnarray}
\text{Res}(\alpha_1)&=&\frac{1}{g^2(\alpha_1 - \alpha_2 )  
 (\alpha_1  - \alpha_3 )   (\alpha_1  -\beta_1  ) }
\nonumber  \\  &\times& \frac{1}{ 
 (\alpha_1 -\beta_2 ) (\alpha_1  - \beta_3 )    } \,,
\nonumber \\
\text{Res}(\alpha_3)&=&\frac{1}{g^2     (  \alpha_3-\alpha_1  ) 
( \alpha_3 -\alpha_2  )(\alpha_3  -\beta_1 ) }
\nonumber  \\  &\times& \frac{1}{ 
 ( \alpha_3-\beta_2 )  (\alpha_3-\beta_3   )    } \,,
\nonumber \\
\text{Res}(\beta_1)&=&\frac{1}{g^2  (\beta_1-\alpha_1   )  
(\beta_1-\alpha_2 ) (\beta_1  -\alpha_3  ) }
\nonumber  \\  &\times& \frac{1}{ 
( \beta_1  - \beta_2 )  (\beta_1  -\beta_3    )    } \,,
\nonumber \\
\text{Res}(\beta_3)&=&\frac{1}{g^2    ( \beta_3-\alpha_1   ) 
 (  \beta_3 -\alpha_2 )    (   \beta_3 -\alpha_3   )     }
\nonumber  \\  &\times& \frac{1}{ 
 (  \beta_3-\beta_1  )  (\beta_3-\beta_2 )  } \,.
\end{eqnarray}
For the expressions above involving a $\beta$,
where a $p_0$ appears, we consider the $\epsilon$
dependence and we compute the discontinuity using the expression 
\begin{eqnarray}
\frac{1}{x\pm i\epsilon}=\mathcal{P}\left(\frac{1}{x}\right)\mp i\pi \delta(x)\,,
\end{eqnarray}
where $\mathcal{P}$ denotes the principal value. For the other terms we just evaluate $\epsilon$
to zero.
Adding all the residues gives
\begin{eqnarray}
&&\text{Disc}\left(  \sum Res \right)=-  \frac{2\pi i}{g^2} \nonumber  \\ &\times&\Big( \frac{\delta 
(p_0-\omega_2 + \overbar \omega_1 )}  {  (\omega_1 - 
\omega_2) (\overbar {\omega}_1 - \overbar \omega_2) 
 (\omega_2 - W) (\overbar \omega_1 - \overbar W)}\nonumber \\&+&
   \frac{\delta (p_0-\omega_1 +\overbar  \omega_2 )   }{(\omega_1
   - \omega_2) (\overbar \omega_1 - 
    \overbar \omega_2)  (\omega_1 - W) (\overbar \omega_2 - \overbar W) }
     \nonumber  \\ &+& \frac{ \delta(p_0+ \overbar \omega_2  - W ) }{ 
       (\overbar \omega_1 -
     \overbar \omega_2) (\omega_1 - W) (\omega_2 - W) (\overbar \omega_2
      - \overbar W) } \nonumber  \\ &+& \frac{ \delta (p_0-\omega_2  + 
    \overbar W ) }{  (\omega_1 - \omega_2)  (\omega_2 - W) (\overbar \omega_1
     - \overbar W) (\overbar \omega_2 - \overbar W)}    \Big)\,,
\end{eqnarray}
where the notation is $\overbar X=X(\vec p-\vec q)$.

Organizing the terms and recalling that we are taking the energy flow in one direction 
where $p_0$ is positive, we drop the first and fourth contribution to arrive at
\begin{eqnarray}
2 \text{Im} \mathcal M&=& \lambda^2 \int \frac{d^{3}
{\vec q}}{(2\pi)^{3}} \frac{2\pi}{g^2}  \left(  \frac{\delta (p_0-\omega_1 +\overbar  \omega_2  )   }{(\omega_1
   - \omega_2) (\overbar \omega_1 - 
    \overbar \omega_2)  (\omega_1 - W)  } \right. \nonumber \\
&\times& \left.  \frac{1}{(\overbar \omega_2 - \overbar W)} +  \frac{ \delta(p_0+ \overbar \omega_2  - W ) }{ 
       (\overbar \omega_1 -
     \overbar \omega_2) (\omega_1 - W) (\omega_2 - W) } \right. \nonumber  \\ &\times&  \left. \frac{1}{ (\overbar \omega_2
      - \overbar W)}    \right) \,,
\end{eqnarray}
where we have used the identity 
$\text{Disc}\mathcal M=2i \text{Im} \mathcal M$.

The second delta is nonvanishing when 
the pole $\alpha_3$ of the first propagator collapses
with the pole $\beta_2$ of the second propagator, as can be seen 
in Fig.~\ref{Fig4}.
In this case both poles pinch the path of integration. 
It is not difficult to show that no other pinching occurs, since the pole $\alpha_2$ has opposite 
sign compared to all other poles and eventually never hits any of them.
 
In order to analyze the pinching condition $\beta_2= \alpha_3$, we consider
  $\xi=e^{i\theta}$ with the angle
   $\theta=\tan^{-1}\left( \frac{\sqrt{4-\beta^2}}{\beta}\right) $
defined  in the interval $0<\theta<\pi $. 

Using these expressions in Eqs.~\eqref{solu} yields
\begin{eqnarray}
\omega_1&=& \frac{1}{3g}\left(1+\cos \frac{\theta}{3} 
 -\sqrt{3}  \sin \frac{\theta}{3}  \right)  \,,
\nonumber \\
\omega_2&=& \frac{1}{3g}\left(1-2\cos \frac{\theta}{3}  \right)     \,,
\nonumber \\
W&=& \frac{1}{3g}\left(1+\cos \frac{\theta}{3}  +\sqrt{3}  \sin \frac{\theta}{3}  \right)   \,.
\end{eqnarray}
We begin with the case where the external space momenta vanish, i.e., $\vec p=0$. In this 
reference frame (center of mass frame), we arrive at 
\begin{eqnarray}\label{pinch}
3gp_0=3\cos \frac{\theta}{3}  +\sqrt{3} \sin \frac{\theta}{3}   \,,
\end{eqnarray}
which has the solution 
\begin{eqnarray}
\theta_0=\frac{\pi}{2} \pm 3 \cos^{-1}\left(\frac{\sqrt{3} gp_0  }{2}    \right)  \,,
\end{eqnarray}
for $1/g<p_0<2/\sqrt{3}g$. Taking the threshold to be
$p_0=1/g$, we have that the pinching occurs at 
$\theta= 0$ ($\Delta=0$)
which lies outside the interval of $\theta$. In this way it is enough to take 
$p_0<1/g$ to avoid the 
propagation of nonphysical degrees of freedom.

Let us consider the case $\vec p \neq 0$. It can be shown that a variation in the energy 
is equivalent to
an increment of the angle $\theta$, which we denote by 
$\delta \theta$. According to Eqs.~\eqref{solu} and~\eqref{def-xi},
 an increment in the momenta is equivalent to an increment of 
second order $\delta \theta \sim g^2$. The new equation at which we arrive is 
\begin{eqnarray}
3gp_0=3a\cos \frac{\theta }{3} +\sqrt{3} b\sin  \frac{\theta}{3}    \,,
\end{eqnarray}
where
\begin{eqnarray}
a=  \frac{1}{3}\left(  1+2 \cos\left(\frac{\delta \theta}{3} 
  \right ) \right )\,,
\end{eqnarray}
and 
\begin{eqnarray}
b=  1-\frac{2}{\sqrt{3}} 
\sin\left(\frac{\delta \theta}{3}     \right)\,.
\end{eqnarray}
It can be seen that, at lowest order in $g$, we arrive at the same result
we have obtained for Eq.~\eqref{pinch}.
In addition, in the region in which $\Delta<0$, we have complex 
solutions and so there is no contribution to the discontinuity.

Finally, the relevant contribution is
\begin{eqnarray}
2 \text{Im} \mathcal M&=& \lambda^2 \int \frac{d^{3}
{\vec q}}{(2\pi)^{3}} \frac{2\pi}{g^2} \nonumber  \\ &\times& \frac{\delta (p_0-\omega_1 +\overbar  \omega_2  )   }{(\omega_1
   - \omega_2) (\omega_1 - W)  (\overbar \omega_1 - 
    \overbar \omega_2)  (\overbar \omega_2 - \overbar W) } \,. 
\end{eqnarray}
Let us define $k_1=q$ and $k_2=q-p$, together with $ \omega_1^{\prime}= \omega_1(\vec k_1)$,
$\omega_2^{\prime}= \omega_2(\vec k_1)$, $ \omega_1''= \omega_1(\vec k_2)$ and
$ \omega_2''= \omega_2(\vec k_2)$, and write
\begin{eqnarray}
&&2 \text{Im} \mathcal M= \lambda^2 \int \frac{d^{4}
{k_1}}{(2\pi)^{4}}  \int \frac{d^{4}
{k_2}}{(2\pi)^{4}}     ( 2\pi)^4     \delta^{(4)}(p-k_1+k_2) \nonumber 
\\  &\times& \left(  \frac{(2\pi)^2 \delta (k^0_1-\omega_1^{\prime} ) \delta (k^0_2- \omega_2 '' )     }{  g(k^0_{1}
   - \omega^{\prime}_2) ( W^{\prime}-k^0_{1}) g(  \omega_1''  -k^0_2)  ( W''-k^0_{2}  )}   \right) \,.
\end{eqnarray}
At this point, it is convenient to define a physical delta 
\begin{eqnarray}\label{delta}
&& \delta^{(\text{phys})}(p^2-m^2 - gp_{0}^{3}) =\sum_{\text{phys},a}
 \frac{\delta(p_0 - p_a)}{|F^{\prime} (p_a)|} \,,
\end{eqnarray}
where $p_a$ are the zeros of the function
$F (p_0)=p_0^2-E_{\vec p}^2  -gp_{0}^{3}$ and where we have to exclude
 the contribution from the 
Lee-Wick-like pole. Considering
\begin{eqnarray}
 \delta^{(\text{phys})}(k_{1}^2-m^2  -gk_{01}^{3}) \theta(k_1^0)&=&
\frac{\delta(k^0_1-\omega_1  )   \theta(k_1^0)}{g(k^0_1-\omega_2) (W-k^0_1)  } \,,
\nonumber \\
 \delta^{(\text{phys})}(k_{2}^2-m^2  -gk_{02}^{3}) \theta(-k_2^0)&=&
\frac{\delta(k^0_2-\omega_2  )\theta(-k_2^0)}{g(\omega_1-k^0_2) (W-k^0_2)  } \nonumber  \,, \\
\end{eqnarray}
where we have used the absolute value in the definition of
Eq.~\eqref{delta}, we rewrite
\begin{eqnarray}
&&2 \text{Im} \mathcal M= \lambda^2 \int \frac{d^{4}
{k_1}}{(2\pi)^{4}}  \int \frac{d^{4}
{k_2}}{(2\pi)^{4}}      (2\pi)^4     \delta^{(4)}(p-k_1+k_2)
\nonumber \\  &\times&  (2\pi)^2 \delta^{(\text{phys})}(k_{1}^2-m^2  -gk_{01}^{3})     
 \delta^{(\text{phys})}(k_{2}^2-m^2  -gk_{02}^{3})  \nonumber \\ &\times& \theta(k_1^0) \theta(-k_2^0)   \,.
\end{eqnarray}
In this way we arrive at 
the phase space sum of the cut diagram, hence, proving
the unitarity constraint 
in our diagram and one-loop unitarity in our theory.

We note that as in the usual case, one could have replaced the propagators
with the physical deltas in the cut diagrams
\begin{eqnarray}
&& \frac{i}{-g(p_0-\omega_1+i\varepsilon) 
 (p_0-\omega_2-i\varepsilon) (p_0-W+i\varepsilon) }  \nonumber   \\  &&\to
  2\pi \delta^{(\text{phys})}(k_{1}^2-m^2  -gk_{01}^{3}) \theta(k_1^0)\,,
\end{eqnarray}
and 
\begin{eqnarray}
&& \frac{i}{-g(q_0-p_0-\omega_1+i\varepsilon) 
 (q_0-p_0-\omega_2-i\varepsilon) (q_0-p_0-W+i\varepsilon) }  \nonumber   \\ && \to 
  \delta^{(\text{phys})}(k_{2}^2-m^2 -gk_{02}^{3}) \theta(-k_2^0) \,,
\end{eqnarray}
simplifying the analysis from the beginning and being of
 potential utility in other models.
\section{Conclusions}\label{con}
In this work we have focused on the Myers-Pospelov effective field theory with dimension-five 
Lorentz violating operators in order to study one-loop unitarity.
In the first part, we have studied the solutions of the dispersion relation for  
purely spacelike, timelike and lightlike backgrounds.
We have found that when $n$ is purely spacelike, one has 
two perturbative solutions which become complex when momenta 
are higher than $\mathcal P$. 
In addition, we have found
that possible issues regarding the canonical quantization may 
arise in highly boosted frames due to spacelike solutions of the dispersion relation.
In the spacelike case, without Lee-Wick particles, we have directly verified the optical theorem 
at one-loop level.
For the the timelike case, we have found two perturbative poles which 
in the limit $g\to 0$ tend to the standard ones, and in accordance with the 
higher time derivative character of the theory, an 
 additional pole corresponding to a particle with negative norm. 
The poles have been characterized according 
to the sign of the discriminant and we have found that 
above the critical energy $\frac{2}{g\sqrt{27}}$ the two poles $\omega_1$ 
and $W$ become complex and move as complex conjugate pairs,
while $\omega_2$ always remains in the real axis.
In this way, we have determined the evolution of the three poles in the complex energy 
plane. The lightlike case is very similar to the timelike case
and presents no new ingredients.

The main part of this investigation has been 
to study whether
 it is possible to preserve unitarity 
by applying 
 the Lee-Wick prescription, which requires us
 to excise  
 the Lee-Wick-like particles
from the physical Hilbert space.
 In particular, we have analyzed the forward scattering of
 antiparticle-particle annihilation with a quartic interaction 
 term. We have studied 
 the bubble diagram with the optical theorem and computed
  the possible 
 contributions to the discontinuity.
It has been found that the Lee-Wick-like pole
 contributes to the discontinuity provided 
a pinching singularity takes place or equivalently when
the path of integration passes between two
  infinitely close poles. Performing a detailed analysis 
  one can show that for real external momenta 
  the pinching condition cannot be fulfilled and so its contribution vanishes.
  
 Finally, by comparing with the cut diagram where we identify the
  sum over intermediate states, conservation of momenta, and on-shell contributions of the propagators,
 we have verified unitarity 
 at the one-loop level. In addition, we have shown that an
  alternative and more direct route may be supplied with a 
 physical delta defined to 
   select only poles associated to stable particles.
  In other words,
 we have shown the equivalence of replacing the propagators on shell
 with physical deltas in the cut diagrams. It may be part of future work to study
  whether this feature is maintained in other Lorentz violating models.
\begin{acknowledgements}
L.B. is supported by DIUFRO through the project: DI16-0075.
C.M.R acknowledges partial support 
by the research project Fondecyt Regular 1140781 and GI172309/C  
UBB.
\end{acknowledgements}


\begin{thebibliography}{99}
  \bibitem{QGV}
V.~A.~Kostelecky and S.~Samuel,
Phys.\ Rev.\  D {\bf 39}, 683 (1989); V.~A.~Kostelecky and R.~Potting,
Nucl.\ Phys.\  B {\bf 359}, 545 (1991); R.~Gambini and J.~Pullin,
  Phys.\ Rev.\ D {\bf 59}, 124021 (1999); J.~Alfaro, 
  H.~A.~Morales-Tecotl and L.~F.~Urrutia,
  Phys.\ Rev.\ Lett.\  {\bf 84}, 2318 (2000); S.~M.~Carroll, 
  J.~A.~Harvey, V.~A.~Kostelecky, C.~D.~Lane and T.~Okamoto,
  Phys.\ Rev.\ Lett.\  {\bf 87}, 141601 (2001).
\bibitem{SME}
D.~Colladay and V.~A.~Kostelecky,
  Phys.\ Rev.\ D {\bf 55}, 6760 (1997); D.~Colladay and V.~A.~Kostelecky,
  Phys.\ Rev.\ D {\bf 58}, 116002 (1998).
\bibitem{SME-nonminimal} 
V.~A.~Kostelecky and M.~Mewes,
Phys.\ Rev.\ D {\bf 80}, 015020 (2009); A.~Kostelecky and M.~Mewes,
Phys.\ Rev.\ D {\bf 85}, 096005 (2012);
A.~Kostelecký and M.~Mewes,
  Phys.\ Rev.\ D {\bf 88}, no. 9, 096006 (2013).
\bibitem{MP}
R.~C.~Myers and M.~Pospelov,
Phys.\ Rev.\ Lett.\  {\bf 90} (2003) 211601.
\bibitem{MP_Limits}
L. Maccione, S. Liberati, A. Celotti and J. G. Kirk,
JCAP {\bf 0710}, (2007) 013; L. Maccione and S. Liberati,
JCAP {\bf 0808}, (2008) 027. 
\bibitem{syn}
R.~Montemayor and L.~F.~Urrutia,
Phys.\ Rev.\  D {\bf 72}, 045018 (2005); 
R. Montemayor and L. F. Urrutia,
Phys. Lett.  {\bf B 606} (2005) 86.
\bibitem{comp} 
V.~A.~Kostelecky and N.~Russell,
  Rev.\ Mod.\ Phys.\  {\bf 83}, 11 (2011).
\bibitem{H-O-C} 
R.\ Casana, M.M.\ Ferreira, R.V.\ Maluf, and F.E.P.\ dos Santos,
Phys.\ Rev.\ D {\bf 86}, 125033 (2012); 
R.\ Casana, M.M.\ Ferreira, Jr., E.O.\ Silva, E.\ Passos, 
and F.E.P.\ dos Santos,
Phys.\ Rev.\ D {\bf 87}, 047701 (2013); 
J.B.\ Araujo, R.\ Casana and M.M.\ Ferreira,
Phys.\ Rev.\ D {\bf 92}, 025049 (2015);
L.~H.~C.~Borges, A.~F.~Ferrari and F.~A.~Barone,
  Eur.\ Phys.\ J.\ C {\bf 76}, no. 11, 599 (2016).
  \bibitem{Rad-Corr}
C.~M.~Reyes, L.~F.~Urrutia and J.~D.~Vergara,
Phys.\ Rev.\  D {\bf 78}, 125011 (2008);  
C.~M.~Reyes, L.~F.~Urrutia and J.~D.~Vergara,
Phys.\ Lett.\  B {\bf 675}, 336 (2009).
\bibitem{dim5}
T.~Mariz,
  Phys.\ Rev.\ D {\bf 83}, 045018 (2011);
 T.~Mariz, J.~R.~Nascimento and A.~Y.~.Petrov,
  Phys.\ Rev.\ D {\bf 85}, 125003 (2012); J.~Leite, T.~Mariz and W.~Serafim,
  J.\ Phys.\ G {\bf 40}, 075003 (2013); T.~Mariz, J.~R.~Nascimento, A.~Y.~Petrov and H.~Belich,
  J. Phys. Communications 1, 045011 (2017).
\bibitem{PU}
A.~Pais and G.~E.~Uhlenbeck, Phys.\ Rev.\  {\bf 79}, 145 (1950).
\bibitem{Lee-Wick}
T.~D.~Lee and G.~C.~Wick,
  Nucl.\ Phys.\ B {\bf 9}, 209 (1969);
 T.~D.~Lee, G.~C.~Wick,
  Phys.\ Rev.\  {\bf D2}, 1033-1048 (1970).
\bibitem{Cut} 
R.~E.~Cutkosky, P.~V.~Landshoff, D.~I.~Olive and J.~C.~Polkinghorne,
Nucl.\ Phys.\ B {\bf 12}, 281 (1969).
\bibitem{G} 
B.~Grinstein, D.~O'Connell and M.~B.~Wise,
Phys.\ Rev.\ D {\bf 77}, 025012 (2008); J.~R.~Espinosa, 
B.~Grinstein, D.~O'Connell and M.~B.~Wise,
Phys.\ Rev.\ D {\bf 77}, 085002 (2008); J.~R.~Espinosa and B.~Grinstein,
Phys.\ Rev.\ D {\bf 83}, 075019 (2011).
\bibitem{Ans-Piva} 
D.~Anselmi,
  JHEP {\bf 1802}, 141 (2018);
  D.~Anselmi and M.~Piva,
  JHEP {\bf 1706}, 066 (2017); D.~Anselmi and M.~Piva,
  Phys.\ Rev.\ D {\bf 96}, no. 4, 045009 (2017).
 \bibitem{unit-LW} 
   C.~M.~Reyes and L.~F.~Urrutia,
  Phys.\ Rev.\ D {\bf 95}, no. 1, 015024 (2017);
  M.~Maniatis and C.~M.~Reyes,
  Phys.\ Rev.\ D {\bf 89}, no. 5, 056009 (2014);
  C.~M.~Reyes,
  Phys.\ Rev.\ D {\bf 87}, no. 12, 125028 (2013);
  J.~Lopez-Sarrion and C.~M.~Reyes,
  Eur.\ Phys.\ J.\ C {\bf 73}, no. 4, 2391 (2013).
\bibitem{unit}
 M.~Schreck,
  Phys.\ Rev.\ D89, no. 10, 105019 (2014);  M.~Schreck,
  Phys.\ Rev.\ D90, no. 8, 085025 (2014).  
\bibitem{dens} 
R.~Potting,
Phys.\ Rev.\ D85, 045033 (2012).
 \bibitem{Ext-leg} 
M.~Cambiaso, R.~Lehnert and R.~Potting,
Phys.\ Rev.\ D90, 065003 (2014); M.~Cambiaso, R.~Lehnert and R.~Potting,
  Phys.\ Rev.\ D {\bf 85}, 085023 (2012).
 \bibitem{ren}  
J.~R.~Nascimento, A.~Y.~Petrov and C.~M.~Reyes,
  Eur.\ Phys.\ J.\ C {\bf 78}, no. 7, 541 (2018); C.~M.~Reyes, S.~Ossandon and C.~Reyes,
  Phys.\ Lett.\ B {\bf 746}, 190 (2015).
 \bibitem{K-R} 
  V.~A.~Kostelecky and R.~Lehnert,
  Phys.\ Rev.\ D {\bf 63}, 065008 (2001).
  \bibitem{Neg-En}  
  D.~Colladay,
  J.\ Phys.\ Conf.\ Ser.\  {\bf 952}, no. 1, 012011 (2018); D.~Colladay, P.~McDonald, J.~P.~Noordmans and R.~Potting,
  Phys.\ Rev.\ D {\bf 95}, no. 2, 025025 (2017); D.~Colladay,
  Phys.\ Lett.\ B {\bf 772}, 694 (2017).
  
 
  
\end{thebibliography}
\end{document}